\documentclass[sigplan,screen,sigconf]{acmart}

\usepackage{multirow}
\usepackage[normalem]{ulem}
\useunder{\uline}{\ul}{}
\def\code#1{\colorbox{backcolour}{\textcolor{blue}{\texttt{#1}}}}
\definecolor{codegray}{rgb}{0.5,0.5,0.5}
\definecolor{backcolour}{rgb}{0.95,0.95,0.92}

\setcopyright{rightsretained}
\acmDOI{}
\acmISBN{}
\acmConference[LATTE '22]{2nd Workshop on Languages, Tools, and Techniques for Accelerator Design}{March 1, 2022}{Lausanne, Switzerland}

\author{Benjamin Biggs}
\affiliation{
  \institution{Imperial College London}
  \country{UK}
}
\author{Ian McInerney}
\affiliation{
  \institution{The University of Manchester}
  \country{UK}
}
\author{Eric C. Kerrigan}
\affiliation{
  \institution{Imperial College London}
  \country{UK}
}
\author{George A. Constantinides}
\affiliation{
  \institution{Imperial College London}
  \country{UK}
}

\begin{document}

\title{High-level Synthesis using the Julia Language}

\begin{abstract}
The growing proliferation of FPGAs and High-level Synthesis (HLS) tools has led to a large interest in designing hardware accelerators for complex operations and algorithms. However, existing HLS toolflows typically require a significant amount of user knowledge or training to be effective in both industrial and research applications. In this paper, we propose using the Julia language as the basis for an HLS tool. The Julia HLS tool aims to decrease the barrier to entry for hardware acceleration by taking advantage of the readability of the Julia language and by allowing the use of the existing large library of standard mathematical functions written in Julia. We present a prototype Julia HLS tool, written in Julia, that transforms Julia code to VHDL. We highlight how features of Julia and its compiler simplified the creation of this tool, and we discuss potential directions for future work.
\end{abstract}


\maketitle

\section{Introduction}


There is a growing need for high throughput, low latency, and energy efficient compute platforms, such as Field Programmable Gate Arrays (FPGAs), by a wide selection of tasks, including machine learning and real-time control.
These platforms often have a high barrier to entry for software developers and researchers who are unfamiliar with digital hardware development, creating what is known as the ``two-language'' problem.
The ``two-language'' problem presents the typical hardware accelerator development workflow as first having a researcher develop and prototype the application algorithms in a high-level language, such as Python or MATLAB, that is simpler to code and is mathematics-friendly.
Then the algorithm is passed to an engineer who translates it into a lower-level language, such as C/C++/VHDL/Verilog, to achieve the desired performance targets.

The Julia language \cite{julia_intro, julia_further_reading} aims to solve this ``two-language'' problem by providing a higher-level and human-readable language that compiles into performant machine code for CPUs.
Extensions to Julia have also been developed to compile normal Julia code to accelerator hardware such as GPUs~\cite{julia_gpu}, distributed processors~\cite{julia_dist_plat}, and TPUs~\cite{fischerAutomaticFullCompilation2018}

The goal of High-level Synthesis (HLS) tools for FPGAs is to solve the ``two-language'' problem by allowing FPGAs to be programmed using higher-level languages instead of VHDL/Verilog.
However, the more mature toolflows like LegUp~\cite{legup_intro} or Vivado only move the accelerator implementation from VHDL/Verilog into languages like C and C++.
Prototype toolflows have been made using higher-level languages, such as Python~\cite{python_hls, quenonHigherLevelSynthesisCodesign2021}, but these usually require the creation of separate compilers/transpilers to translate the chosen language into an FPGA design.

In this paper, we propose to use Julia as a high-level language for FPGA accelerators, and that the HLS toolflow should be written in Julia as well.
We start in Section~\ref{sec:advantages} by describing the features Julia has that makes it an ideal language for FPGA HLS, and present a brief report on a prototype Julia HLS tool in Sections~\ref{sec:synthesis} and~\ref{sec:evaluation}.

\section{Advantages of the Julia Language}
\label{sec:advantages}

Julia was developed as a language for scientific computing that sought to provide both ease of programming and performant low-level machine code.
The Julia ecosystem contains many mathematical packages, and unlike other higher-level languages, many of these packages are written in pure Julia instead of being a thin wrapper over an underlying C/C++ library.
This means a Julia HLS toolflow could recurse into the packages and generate hardware for them using the HLS compiler instead of having to provide and substitute in hand-written function blocks.

The Julia compiler is also mainly written in Julia, and provides several places to hook into the compilation flow to modify/observe the compilation.
This means that an HLS toolflow for Julia that is written in Julia can benefit by reusing existing compiler stages and optimisation passes used for the CPU compiler.
This also allows for more easier CPU simulation of the desired algorithm, since the initial stages of the compiler flow can be shared between the CPU and HLS tools.
Three other design features of Julia that provide benefits for an HLS toolflow are \textit{built-in data-flow type inference}, \textit{multiple dispatch for method specialisation}, and \textit{extensive metaprogramming features}.

\subsection{Data-flow type inference}
In Julia, explicit type declarations of variables are not required for the code to successfully run because Julia executes a data-flow type inference pass in its compiler to infer these types.
 The type inference pass tries to produce ``type-stable'' code, where all the types in the code are known explicitly at compile time, by combining any known type information on a variable alongside a type lattice to propagate types to every variable.
 An HLS toolflow can then utilize the typed Intermediate Representation (IR) after the type inference compiler pass when generating the hardware.

\subsection{Multiple dispatch}
Julia utilizes multiple dispatch, where the compiler and/or run-time environment examines the types of all function inputs, and then chooses a version of the function that has been written for the specified input types.
 Using multiple dispatch allows the creation of more efficient and readable code by allowing developers to write functions that contain the generic operations, while also providing functions optimised for the provided input types.
 Multiple dispatch can also be used for built-in operators, like ``+'', allowing the HLS toolflow to provide FPGA-specific methods for the basic operators that can be automatically chosen by the Julia compiler when using the HLS toolflow, eliminating the need to modify the actual application source code to reference the HLS-specific basic operators instead of the normal CPU operators.

\subsection{Meta-programming}
Julia has an extensive metaprogramming capability provided in the language, allowing Julia code to modify and transform other Julia code before it is compiled.
 This is accomplished by writing Julia macros, which can be attached to variables, functions and control flow statements.
 Macros have direct access to the Abstract Syntax Tree (AST) for the associated code blocks and are evaluated before the lowered IR is generated.
 The Julia AST includes information such as function calls, loops and conditional statements, allowing the macros to operate directly on the control flow operators.
 HLS analysis and optimization passes can be implemented using Julia macros to modify the AST and interact with the control flow directly, instead of having to infer the original control flow from the LLVM basic blocks and phi-nodes.
 
\begin{figure}[tb!]
    \centering
    \includegraphics[width=0.45\textwidth]{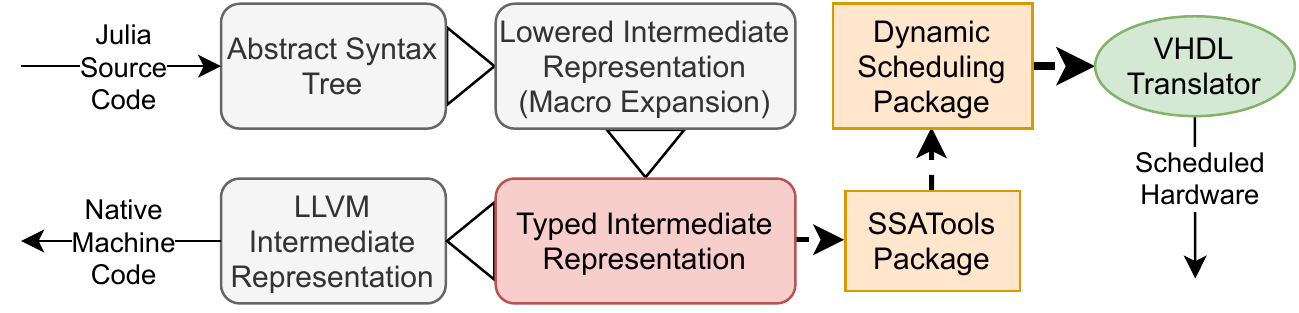}
    \caption{Overview of the Julia compiler \& HLS flow.}
    \label{compiler_flow}
\vspace*{-6mm}
\end{figure}

\section{Transforming Julia into Hardware}
\label{sec:synthesis}
The Julia HLS toolflow takes a subset of the Julia source code and transforms it into synthesisable HDL that can be tested in simulation.
The flow is contained inside two Julia packages, SSATools.jl and DynamicScheduling.jl, that convert the Julia code into a Control Data-Flow Graph (CDFG) made from elastic components detailed in the Dynamatics tool~\cite{dynamatics_p} which then generates VHDL for synthesis with Vivado HLS, as shown in Figure~\ref{compiler_flow}.

\subsection{Initial Experiments} 
The dynamic scheduling in Dynamatics formed the baseline for the initial results of the toolflow.
 The testcases used were chosen to test simple programmatic constructs like conditional statements and looping as well as the decomposition of mathematical functions by the Julia compiler.

 The first testcase was an if-else conditional with three return statements based on the computation of the two input variables. This was chosen to test the processing of branching in the SSATools construction of the CDFG.
 Another testcase was a loop implementation of the power function, where the first input is raised to the power of the second through repeated multiplication in a while loop.
 The Julia compiler decomposes the loop into several basic blocks as part of the typed IR, and the CDFG generation determines the dependencies of the variables within those blocks.
 The final testcase implemented the Newton-Raphson method for determining the roots of a simple polynomial.
 This combined mathematical operations, conditionals and looping to fully explore the current version of the toolflow. 

The results of these testcases are outlined in Table~\ref{tab:synth}, and are compared to the Dynamatics baseline for the same testcases implemented in C++.
 The number of components directly corresponds to FPGA resource usage, since the programs are dynamically scheduled.

\begin{table}[tb!]
\centering
\begin{tabular}{c|cc|cc}
\multirow{2}{*}{Program} & \multicolumn{2}{c|}{Basic Blocks} & \multicolumn{2}{c}{Components} \\
                         & Julia            & C++            & Julia           & C++ \\ \hline 
\code{if\_else}                 & 5                & 3              & 41              & 32  \\
\code{power}                    & 4                & 4              & 61              & 64  \\
\code{newton\_raphson}          & 10               & 6              & 225             & 147 
\end{tabular}
\caption{Synthesis results for the Julia HLS toolflow.}
\label{tab:synth}
\vspace*{-11mm}
\end{table}

\section{Evaluation and Future Work}
\label{sec:evaluation}

The difference in resources between the Julia HLS toolflow and Dynamatics is due to the number of basic blocks generated for the IR.
 The Dynamatics flow initially had more basic blocks in the LLVM IR in each case, but the LLVM optimization passes consolidated blocks together.
 Equivalent versions of these passes should be created for the Julia flow.
 The main limitation of the current toolflow is the lack of data memory integration.
 Adding memory support requires the extraction of additional information from the typed IR, as well as component support for handling memory accesses on the FPGA.
 Additionally, static scheduling for the operations should be implemented to allow for more resource-efficient designs.
 Finally, Julia is designed to be an open-source language, so the HDL output of our tool should be usable in open-source tools like Yosys~\cite{yosys} as well as closed-source tools like Vivado. Any IP cores required should be generated using FloPoCo~\cite{flopoco} or other open-source HDL libraries.

\section{Conclusion}
The aim of this work was to detail the key features of the Julia
language that make it a powerful and adaptable language for the
generation of custom hardware designs for FPGAs.
 The higher-level nature of the language makes complex mathematical functions and algorithms easier to represent while also generating optimal low-level machine code.
 We have discussed several of the benefits the language provides, and showed our first attempt at designing a Julia-based toolflow that is able to translate a subset of the Julia language's typed IR into a CDFG for VHDL code generation.
 This work forms the basis for an array of future work, including the implementation of memory support and alternative scheduling methods.

\bibliography{bib}

\end{document}